\documentclass[9pt]{article}

\usepackage{bm}
\usepackage[letterpaper,hmargin=1in,vmargin=1in]{geometry}
\usepackage{times}
\usepackage{multirow}
\usepackage{dcolumn}
\usepackage{tabularx}
\usepackage{graphicx}
\usepackage{subfigure}
\usepackage{ae}
\usepackage{lettrine}
\usepackage{color}
\usepackage{amsmath,amsthm,amssymb,amsfonts,boxedminipage}
\usepackage{epstopdf}
\newcommand\address[1]{\hskip2.25pc \parbox{.8\textwidth}{ \noindent%
   \footnotesize \it \begin{center} #1 \end{center}\rm }  \normalsize \vskip-.2cm }

\renewcommand\title[1]{\bf \hskip2.25pc \parbox{.8\textwidth}{ \noindent%
   \LARGE \bf \begin{center} #1 \end{center} \rm } \vskip.1in \rm\normalsize }

\renewcommand\author[1]{\hskip2.25pc \parbox{.8\textwidth}{ \noindent%
   \normalsize \bf \begin{center} #1 \end{center}\rm } \vskip-1pc }
\makeatletter
\thm@headfont{\sc}
\makeatother

\makeatletter
\renewcommand\@biblabel[1]{#1.}
\renewcommand{\thesubfigure}

\begin{document}

\title{Quantum restricted Boltzmann machine is universal for quantum computation}
\author{Yusen Wu$^{1,2,4}$, Chunyan Wei$^{1,3}$, \\Sujuan Qin$^{1}$, Qiaoyan Wen$^{1}$, and $^{*}$Fei Gao$^{1,4}$ }
\address{$^{1}$State Key Laboratory of Networking and Switching Technology,
Beijing University of Posts and Telecommunications, Beijing, 100876, China\\
$^{2}$State Key Laboratory of Cryptology, P.O. Box 5159, Beijing, 100878, China\\
$^{3}$School of Mathematical Science, Luoyang Normal University, Luoyang 471934, China\\
$^{4}$Center for Quantum Computing, Peng Cheng Laboratory, Shenzhen 518055, China\\
$^\ast$gaof@bupt.edu.cn}

\begin{quote}
The challenge posed by the many-body problem in quantum physics originates from the difficulty of describing the nontrivial correlations encoded in the many-body wave functions with high complexity. Quantum neural network provides a powerful tool to represent the large-scale wave function, which has aroused widespread concern in the quantum superiority era. A significant open problem is what exactly the representational power boundary of the single-layer quantum neural network is. In this paper, we design a 2-local Hamiltonian and then give a kind of Quantum Restricted Boltzmann Machine (QRBM, i.e. single-layer quantum neural network) based on it. The proposed QRBM has the following two salient features. (1) It is proved universal for implementing quantum computation tasks. (2) It can be efficiently implemented on the Noisy Intermediate-Scale Quantum (NISQ) devices. We successfully utilize the proposed QRBM to compute the wave functions for the notable cases of physical interest including the ground state as well as the Gibbs state (thermal state) of molecules on the superconducting quantum chip. The experimental results illustrate the proposed QRBM can compute the above wave functions with an acceptable error.


\date{\today}
\end{quote}
\quad The wave function is an important object in quantum physics and is difficult to be characterized in the classical world. Actually, the  wave function encodes all of the information of a complex molecule on a quantum state which needs an extremely large-scale space to characterize. Generally, the scale of the space increases exponentially with the size of the physical system, therefore a large-scale wave function with complex correlations among the subsystems needs too enormous resources to depict for a classical computer. To address the above problem, Feynman proposed the idea that one can utilize a quantum computer to simulate complex wave functions \cite{Feyman}, then some quantum algorithms for solving many-body problems of interacting fermions were proposed \cite{Abrams1997Simulation,Aspuru2005Simulation}. These algorithms start from a ``good'' initial state that has a large overlap with the target state \cite{Kandala2018QVE}. Afterwards, they perform the phase estimation algorithm onto the initial state to encode the eigen-values of the Hamiltonian into the quantum register. Noting that though these algorithms can produce an extremely accurate energy for solving quantum chemistry and quantum material problems, they apply stringent requirements on the coherence of the quantum hardware devices which are inaccessible with current technology.

To reduce the coherence requirements on the quantum devices, classical-quantum hybrid algorithms were delivered. This kind of algorithms involve minimizing a cost function that depends on the parameters of a quantum gate sequence. Cost evaluation occurs on the quantum computer, with speed-up over classical evaluation, and the classical computer utilizes this cost information to adjust the parameters of the ansatz with the help of suitable classical optimization algorithms. As one of the most representative classical-quantum hybrid algorithms, the Variational Quantum Eigensolver (VQE) utilizes Ritz's variational principle to prepare approximations to the ground state and its energy \cite{MYung2014QVE}. However, the efficiency of VQEs is limited by the number of parameters that scales quartically with the number of spin orbitals that are considered in the single- and double- excitation approximation. To improve the VQE algorithm, the hardware-efficient trial states were introduced, which are composed by the single-qubit Euler rotation part and the entanglement part \cite{Kandala2018QVE}, and the hardware-efficient ansatz can be efficiently implemented on the Noisy Intermediate-Scale Quantum (NISQ) devices.

The quantum neural network is a significant ansatz in simulating many-body systems \cite{Carleo2017QRBM}. In fact, the neural network is a powerful tool to interpret complex correlations in multiple-variable functions or probability distributions in the classical world. Numerical experiment suggests that the single layer neural network, i.e. the Restricted Boltzmann Machine (RBM), provides a good solution to several many-body systems, such as the transverse-field Ising model and the antiferromagnetic Heisenberg model \cite{Carleo2017QRBM}. However, the representation power of the RBM is not sufficient for implementing the universal quantum computation tasks. Duan et al. \cite{Xun2017DNN} analyzed the representational power of the RBM, and indicated that the RBM cannot characterize some of the quantum states, such as the projected entangled pair states and quantum enhanced feature states \cite{NISQSVM, Xun2017DNN}.

Afterwards, researchers proposed the Quantum Restricted Boltzmann Machines (QRBM) to efficiently simulate some many-body systems. In 2018, Xia et al. \cite{RongQMLESC2018} introduced a series of single-qubit rotation to construct a marginal state of the QRBM, which provides a good solution on simulating the Hydrogen molecule as well as the Water molecule. Zhang et al. \cite{Shengyu2019UCRBM} proposed a variational quantum algorithm to efficiently train the QRBM, where the proposed algorithm reduced the required ancillary qubits. Recently, Carleo et al. \cite{Carleo2020QRBM} presented an extension of quantum neural network to model interacting fermionic problems, and Kerstin et al. \cite{Kerstin2020DQNN} indicated that the \textbf{Deep QRBM} can implement the universal quantum computation tasks. The previous works show outstanding performance in some notable cases of physical interest that are difficult for classical RBM. These results suggest the QRBM has stronger representational power compared with the RBM, and this conjecture is also verified in some quantum machine learning algorithms \cite{NISQSVM, Harrow2009Quantum, Yu2019Quantum, Yu2017Quantum, lin2018Toeplitz, Wu2019QCRF, Wittek2014Quantum}. However, as pointed out by Roger G. Melko et al. \cite{Roger2019RBM}, whether the QRBM can implement universal quantum computation tasks is still a significant open problem.


In this paper, we utilize a 2-local Hamiltonian to induce a kind of QRBM, which we call 2-Local QRBM (2L-QRBM). Different from the previous QRBM, the 2L-QRBM has connections between visible nodes. Specifically, our model has two salient features. (1) It is proved universal for implementing quantum computation tasks. To do this, we consider the simplest case for the 2L-QRBM with only $1$ hidden nodes $(M=1)$. We provide three theorems to construct a map between the 2L-QRBM and the quantum circuit model. Given an arbitrary quantum state $|\alpha\rangle$ that is produced by a quantum circuit, the proof begins at indicating that the state $|\alpha\rangle$ can be encoded as the ground state of a 2-local Hamiltonian $\mathcal{H}$ (Theorem 2). Then we propose how to construct a 2L-QRBM whose corresponding trial state $|\Phi\rangle$ is $\mathcal{O}(\epsilon)$ close to the ground state of $\mathcal{H}$ (Theorem 3, 4), where $\epsilon$ is the approximation error. It implies that, compared with the classical RBM which cannot simulate an arbitrary quantum state \cite{Xun2017DNN}, QRBM illustrates the quantum advantages in terms of the representative power. (2) The proposed 2L-QRBM can be efficiently transformed to a quantum circuit and then be easily implemented on the NISQ devices. Based on this advantage, we validate the accuracy of the proposed 2L-QRBM by studying the Hydrogen molecule as well as the Water molecule on the quantum simulator. And we also utilize 2L-QRBM to compute the Gibbs states of Haldane chains on a superconducting device. The power of the 2L-QRBM is demonstrated, obtaining state-of-art accuracy in computing ground states and Gibbs states.
\section*{Results}
\noindent \textbf{The construction of 2L-QRBM.} 
The definitions of RBM and QRBM refer to Methods. Here, to satisfy the representation power for implementing universal computation tasks, we design a $2^{(N+M)}\times2^{(N+M)}$ bipartite Hamiltonian:
\begin{eqnarray}
\begin{split}
\mathcal{H}_{RBM}(\bm{\theta})=\sum\limits_{i=1}^{N}\sum\limits_{t\in\{x,y,z\}}b_i^tv_i^t+\sum\limits_{j=1}^{M}m_jh^{z}_{j}+\sum\limits_{i=1}^N\sum\limits_{j=1}^{M} W_{ij}v_i^zh_j^z+\sum\limits_{s=1}^{N-1}\sum\limits_{k=s+1}^N\sum\limits_{t\in\{x,y,z\}}K^{t}_{sk}v_s^tv_k^t,
\end{split}
\end{eqnarray}
which induces a quantum Boltzmann machine constituted by one layer of $N$ nodes $\mathbf{v}=\{v_i\}_{i=1}^N$ and a single hidden layer of $M$ auxiliary nodes $\mathbf{h}=\{h_j\}_{j=1}^M$ (see Fig.1). The notation $v_i^t$ represents the Pauli operator $\sigma_i^t (t\in\{x,y,z\})$ defined on the $i$-th visible node, $h_j^z$ denotes the Pauli operator $\sigma_j^z$ on the $j$-th hidden node, and $\bm{\theta}=\{b_i^t,m_j,W_{ij},K_{sk}^t\}$ is real-valued Boltzmann parameter. In the Hamiltonian $\mathcal{H}_{RBM}(\bm{\theta})$, the first two terms indicate the energy operators defined on the visible qubits and hidden qubits, respectively. The third term represents the connections between the visible layer and the hidden layer, and the final term expresses the intersections between visible nodes. As this quantum Boltzmann machine is induced by a 2-local $\mathcal{H}_{RBM}(\bm{\theta})$ without involving the interaction terms $h_i^z\otimes h_j^z$ between the hidden nodes, it can be called 2-Local Quantum Restricted Boltzmann Machine \cite{Kulchytskyy2016Quantum}.

\begin{figure}
  \begin{center}
  \includegraphics[width=0.55\textwidth]{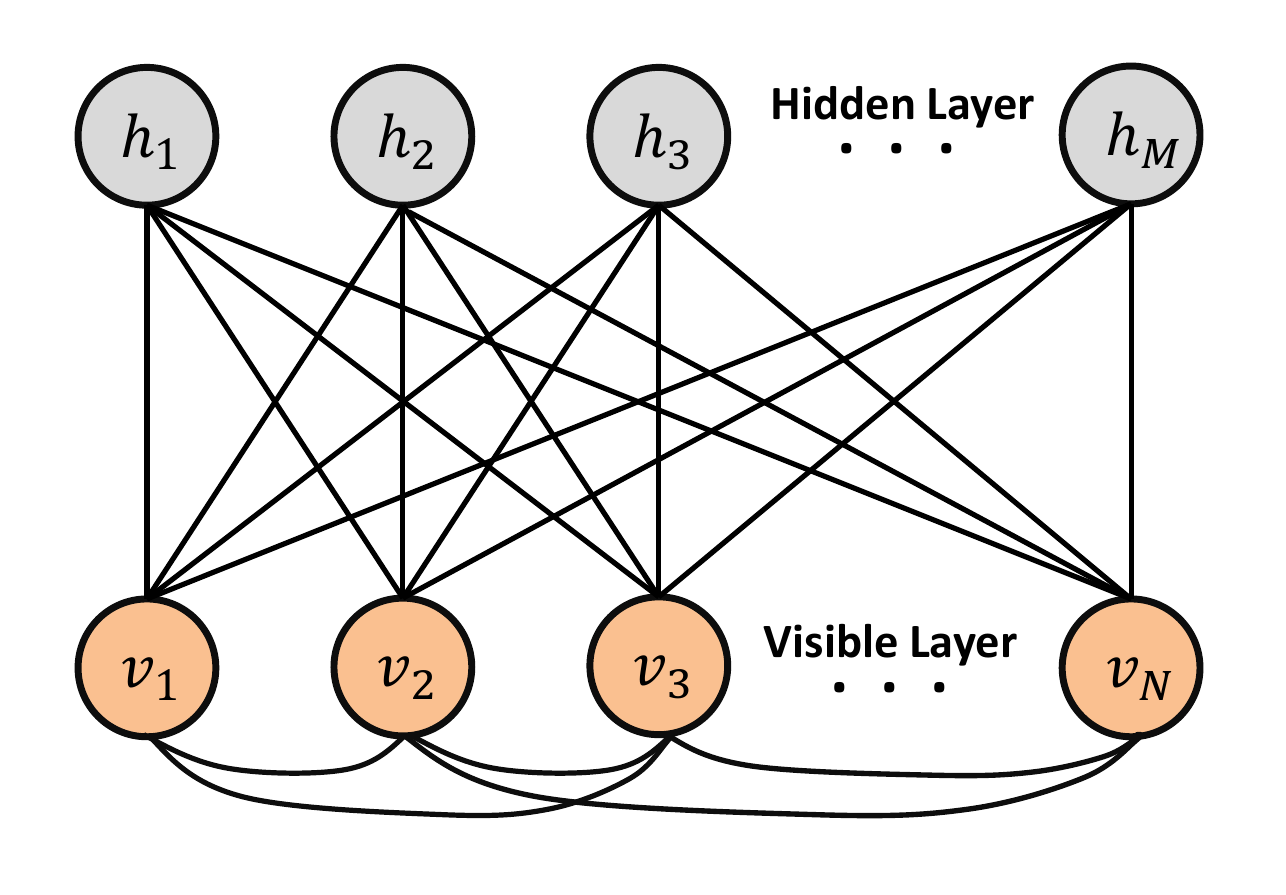}\\
  \caption{The 2L-QRBM induced by the Hamiltonian $\mathcal{H}_{RBM}(\bm{\theta})$. It has $N$ visible nodes (orange circle) as well as $M$ hidden nodes (grey circle), and black solid lines indicate the coupling relationship between different nodes.}
  \end{center}
\end{figure}

The trial state of 2L-QRBM can be created with a two-step approach. First, entangle $N+M$ qubits (including all visible and hidden nodes) according to
\begin{align}
|\Psi_{vh}(\bm{\theta})\rangle=\frac{e^{\mathcal{H}_{RBM}(\bm{\theta})}H^{\otimes (N+M)}|0\rangle^{\otimes (N+M)}_{vh}}{\sqrt{\langle+|^{\otimes (N+M)}e^{2\mathcal{H}_{RBM}(\bm{\theta})}|+\rangle^{\otimes (N+M)}}},
\end{align}
where $H$ is the Hadamard gate, $|+\rangle=\frac{1}{\sqrt{2}}(|0\rangle+|1\rangle)$, and the denominator is a normalization factor. Note that $e^{\mathcal{H}_{RBM}(\bm{\theta})}$ is a non-unitary operator that is difficult to implement on the quantum computer in general.
To solve this problem, we will propose a method to transform $e^{\mathcal{H}_{RBM}(\bm{\theta})}$ into a series of fundamental quantum gates so that one can implement it efficiently on the NISQ devices.

Second, once the wave function $|\Psi_{vh}(\bm{\theta})\rangle$ is generated, all the hidden nodes (qubits) will be post-measured by $|+\rangle$. The measurement should be executed several times until all the hidden nodes (qubits) are projected onto the state $|+\rangle$. After that, the 2L-QRBM trial state can be expressed as
\begin{align}
|\Psi_v(\bm{\theta})\rangle=\frac{\langle+|_{h}^{\otimes M}|\Psi_{vh}(\bm{\theta})\rangle}{\sqrt{\langle\Psi_{vh}(\bm{\theta})|P_{+}^{(h)}|\Psi_{vh}(\bm{\theta})\rangle}}=\frac{1}{N_{\mathbf{v}}}\sum_{\mathbf{h}}e^{\mathcal{H}_{RBM}(\bm{\theta},\mathbf{h})}|+\rangle^{\otimes N},
\label{Eq:2LQRBM}
\end{align}
in which $P_{+}^{(h)}=(|+\rangle\langle+|)_1\otimes...\otimes(|+\rangle\langle+|)_M$ is the measurement operator, and $N_{\mathbf{v}}$ is the normalization factor. $\mathcal{H}_{RBM}(\bm{\theta},\mathbf{h})=\sum_{i=1}^{N}\sum_{t\in\{x,y,z\}}b_i^tv_i^t+\sum_{j=1}^{M}m_j(1-2h_j)I_j+\sum_{i=1}^N\sum_{j=1}^{M} W_{ij}v_i^z(1-2h_j)+\sum_{s=1}^{N-1}\sum_{k=s+1}^N\sum_{t\in\{x,y,z\}}K^{t}_{sk}v_s^tv_k^t$ is a operator acting on the $N$ visible qubits.
One can utilize the trial state $|\Psi_v(\bm{\theta})\rangle$ to approximate the target wave function of the realistic physical system.

\noindent\textbf{2L-QRBM is universal for quantum computation.}
It is well known that the quantum circuit model is universal for quantum computation task, that is, there exists sets of gates acting on a constant number of qubits that can efficiently simulate a quantum Turing machine \cite{Jacob2008}.
In 2005, Aharonov et al. \cite{Aharonov2005AQC} proved that the Adiabatic Quantum Computation (AQC) is also universal for quantum computation, that is, the AQC can simulate the output of any quantum circuit in the polynomial time. Here, we prove that the 2L-QRBM is universal for implementing quantum computation tasks in a similar way.

\textbf{Theorem 1.} \emph{The 2L-QRBM induced by the Hamiltonian $\mathcal{H}_{RBM}(\bm{\theta})$ (Eq.(1)) can implement universal quantum computation tasks. That is, for an arbitrary quantum circuit whose output is denoted as $|\alpha\rangle$ and an arbitrary positive value $\epsilon$, there exists a 2L-QRBM trial state $|\Psi_v(\bm{\theta})\rangle$ that is $\mathcal{O}(\epsilon)$ close to $|\alpha\rangle$.}

This result can be induced immediately from the Theorems 2-4 in Methods, and we here only give a brief overview. Detailed proof refers to Methods. We first indicate that the output state $|\alpha\rangle$ of
an arbitrary quantum circuit can be approximated by the ground state of a 2-local Hamiltonian $\mathcal{H}$ (see Theorem 2 in Methods). Furthermore, to reduce the variational parameters, we prove that the ground state of any 2-local Hamiltonian $\mathcal{H}$ can be approximated by that of a simplified 2-local Hamiltonian which is in the form of $\mathcal{H}_{s}\bm{(\widetilde{\theta})}=\sum_{i=1}^{N}\sum_{t\in\{x,y,z\}}b_i^tv_i^t+\sum_{s=1}^{N-1}\sum_{k=s+1}^N\sum_{t\in\{x,y,z\}}K^{t}_{sk}v_s^tv_k^t$, where $\bm{\widetilde{\theta}}=\{b^t_i,K^t_{sk}\}$. This approximation successfully truncates nearly half of the variational parameters compared with the general 2-local Hamiltonian (see Theorem 3 in Methods).
Finally, we prove that there exists a 2L-QRBM trial state converging towards the ground state of $\mathcal{H}_{s}\bm{(\widetilde{\theta})}$ by introducing a positive `phase shift' $\lambda^*$ (see Theorem 4 in Methods). We consider the simplest case of 2L-QRBM, that is, $M=1$ and $m_j=0$. Given an arbitrary small positive value $\epsilon$, the 2L-QRBM trial state $|\Psi_v(\bm{\theta}^*)\rangle$ with Boltzmann parameters $\bm{\theta}^*=\{m_j,W_{i1},b_i^t,K^t_{sk}\}=\{0, \ln(e^{\lambda^*\tau/N}+\sqrt{e^{2\lambda^*\tau/N}-1}), -\tau f(\bm{\widetilde{\theta}})\}$ is $\mathcal{O}(\epsilon)$ close to the ground state of the simplified Hamiltonian $\mathcal{H}_{s}\bm{(\widetilde{\theta})}$ (simply denoted as $\mathcal{H}_{s}\bm{(\widetilde{\theta})}=\sum_j\widetilde{\theta_j}P_j, P_j\in\{v_i^t,v_s^tv_k^t\}, t\in\{x,y,z\}$), where the time evolution parameter $\tau=\mathcal{O}(\textrm{poly}(1/\epsilon,N))$ and $f(\widetilde{\theta_j})=\widetilde{\theta_j}-\frac{(E_0+\delta)}{2^N}\sum_k\mathrm{Tr}\left(P_j|\psi_k\rangle\langle\psi_k|\right)$.
The quantum state $|\psi_k\rangle$ is the $k$-th excited state of $\mathcal{H}_{s}\bm{(\widetilde{\theta})}$, $\widetilde{\theta_j}$ is the $j$-th component of $\bm{\widetilde{\theta}}$
and $\delta$ is a small positive value that is smaller than the spectral gap of $\mathcal{H}_{s}\bm{(\widetilde{\theta})}$ (see Methods). Thus, the output state $|\alpha\rangle$ of
an arbitrary quantum circuit can be efficiently approximated by a 2L-QRBM trial state.

Theorem 1 indicates that our 2L-QRBM model is complete for the description of many-body quantum system. As shown in \cite{Xun2017DNN}, the classical analogue of the 2L-QRBM, is incapable of this kind of task. Then what kind of advantage our 2L-QRBM has in the description of wave function?


\noindent \textbf{Quantum advantages of 2L-QRBM.} In the classical RBM, the wave function $|\Psi(\bm{\theta})\rangle=\sum_{\mathbf{v}}\Psi_{\mathbf{v}}(\bm{\theta})|\mathbf{v}\rangle$, in which the amplitude $\Psi_{\mathbf{v}}(\bm{\theta})=\sum_{\mathbf{h}}e^{-E_{\bm{\theta}}(\textbf{v,h})}$, and the energy function $E(\mathbf{v},\mathbf{h})=\sum_{i}b_iv_i+\sum_{j}m_jh_j+\sum_{ij}W_{ij}v_ih_j+\sum_{i,j}K_{ij}v_iv_j$. Note that the amplitude (marginal distribution) $\Psi(\mathbf{v})$ can be computed as
\begin{align}
\Psi(\mathbf{v})=\exp\left(\sum_{i}b_iv_i+\sum_{i,j}K_{ij}v_iv_j\right)\prod\limits_{j=1}^M\cosh\left(m_j+\sum\limits_iW_{ij}v_i\right),
\end{align}
which means that $\Psi(\mathbf{v})$ can be calculated in polynomial time under given input values of $\mathbf{v}=(v_1v_2...v_N)$. It is exactly the property which limits the performance of classical RBM. On one hand, if a quantum state has the classical RBM representation, the computation complexity of computing $\Psi(\mathbf{v})$ is of class $\textbf{P/}\textbf{poly}$, that is, this problem can be solved by a polynomial-size circuit even if the circuit cannot be constructed efficiently in general. On the other hand, Duan et al. \cite{Xun2017DNN} proved that simulating some kind of quantum states, such as $|\Phi(\mathbf{x})\rangle=\exp\left(i\sum_{S\subset [m]}\phi_S(\mathbf{x})\prod_{i\in S}\sigma_i^z\right)|0\rangle^m$, projected entangled pair state and the ground state of gapped Hamiltonians, is $\textbf{\#P}$-hard for classical computer. Therefore, these states cannot be efficiently simulated by the classical RBM, otherwise $\textbf{\#P}\subset \textbf{P/\textrm{poly}}$  will be induced, which obviously means that polynomial hierarchy (\textbf{PH}) collapses. Luckily, our 2L-QRBM has significant quantum advantage in simulating these states. Actually, as Theorem 1 shows, 2L-QRBM is complete for the description of wave functions.
\begin{figure}[htbp]
\centering
  \centering
  \subfigure[(a)]{
  \includegraphics[width=0.48\textwidth]{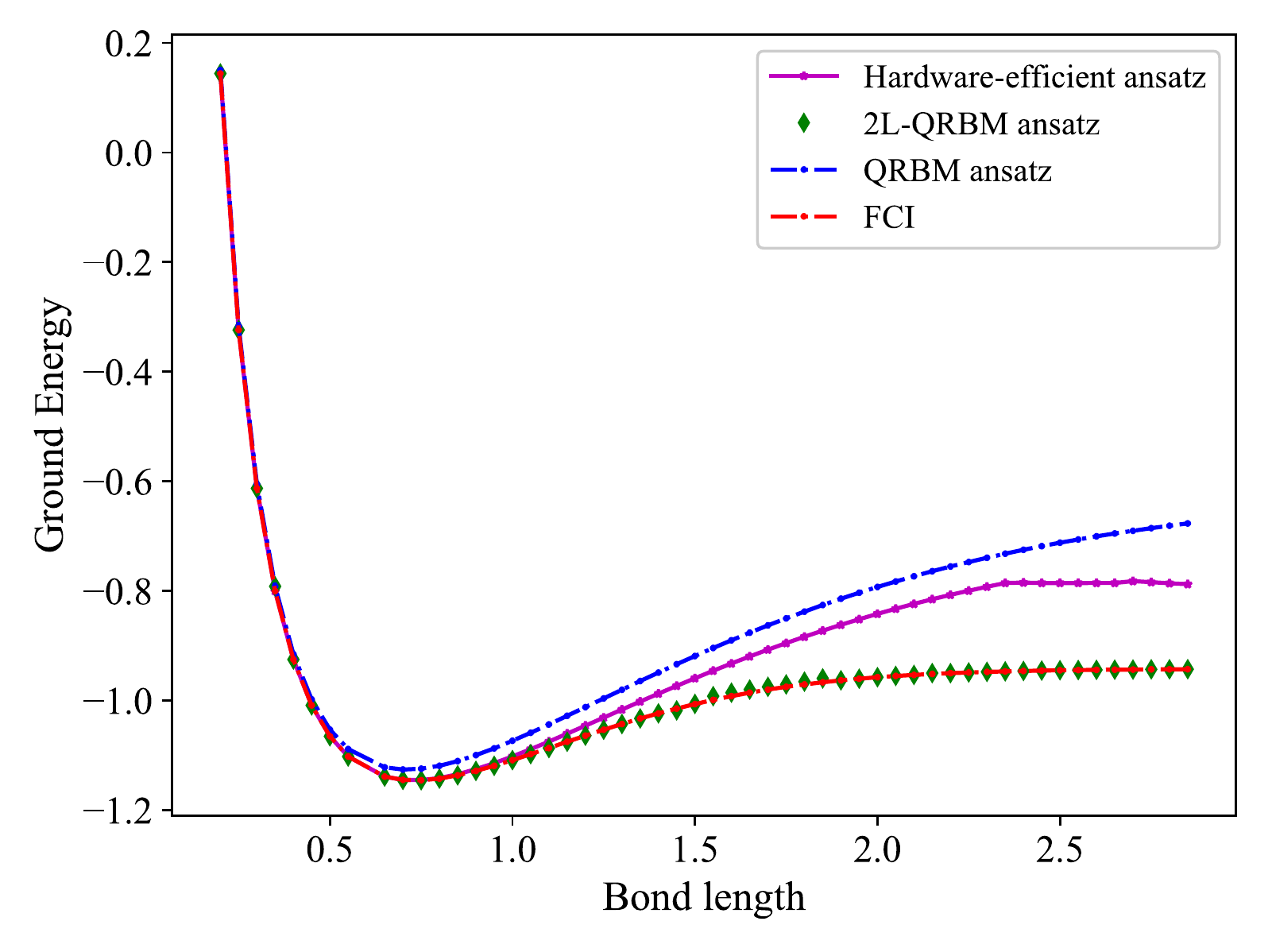}
  }
  \subfigure[(b)]{
  \includegraphics[width=0.48\textwidth]{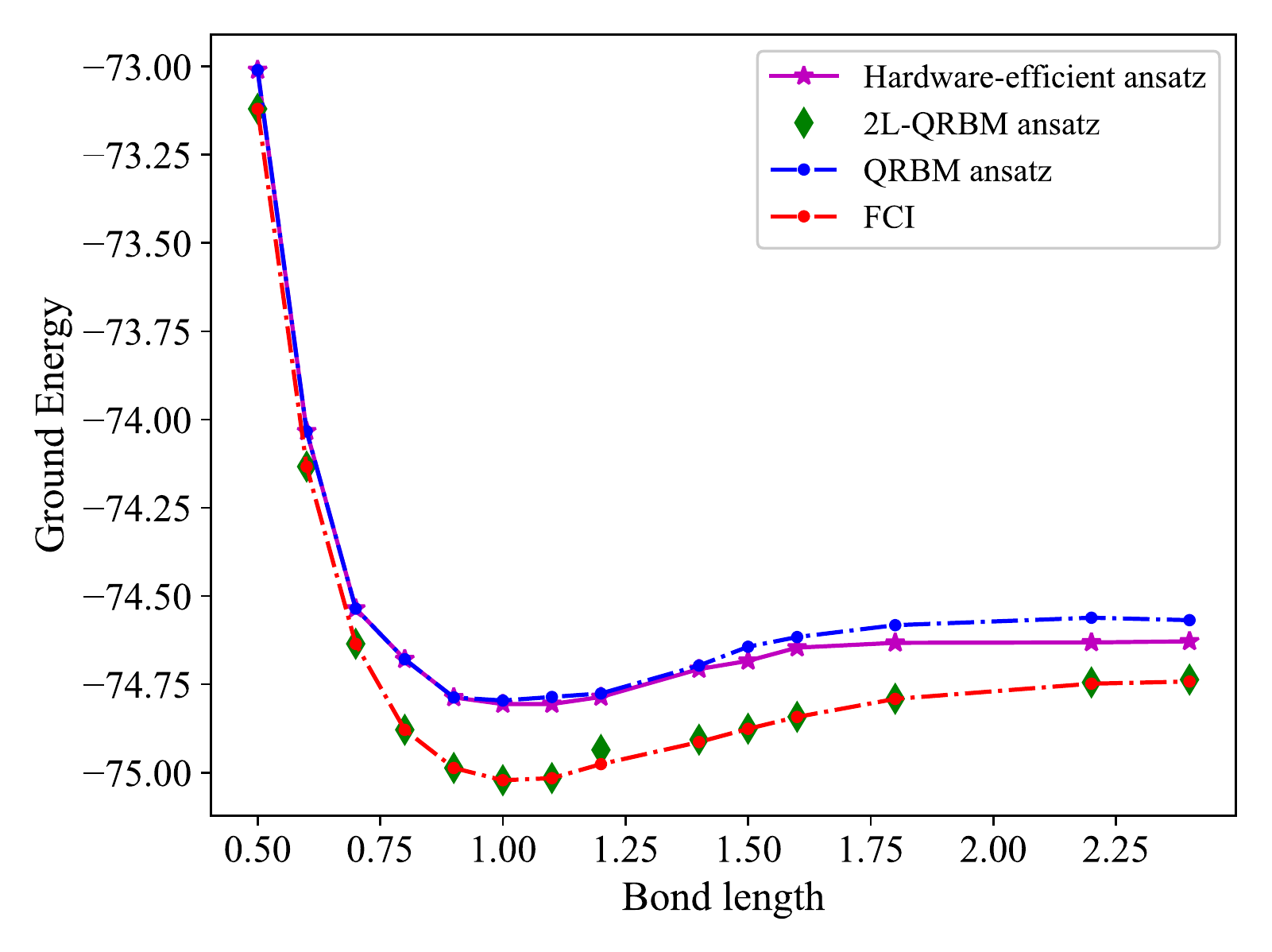}
  }
  \DeclareGraphicsExtensions.
  \caption{Bond dissociation curves of the Hydrogen molecule (a) and the Water molecule (b). The curves are obtained by repeated computation of the ground state energy for several bond length values. The simulation results are computed by the ProjectQ \cite{ProjectQ} (We choose the swap operator as the entanglement operator in the hardware-efficient ansatz).}
  \label{Fig:GroundEnergy}
\end{figure}
\begin{figure}[htbp]
\centering
  \centering
  \subfigure[(a)]{
  \includegraphics[width=0.48\textwidth]{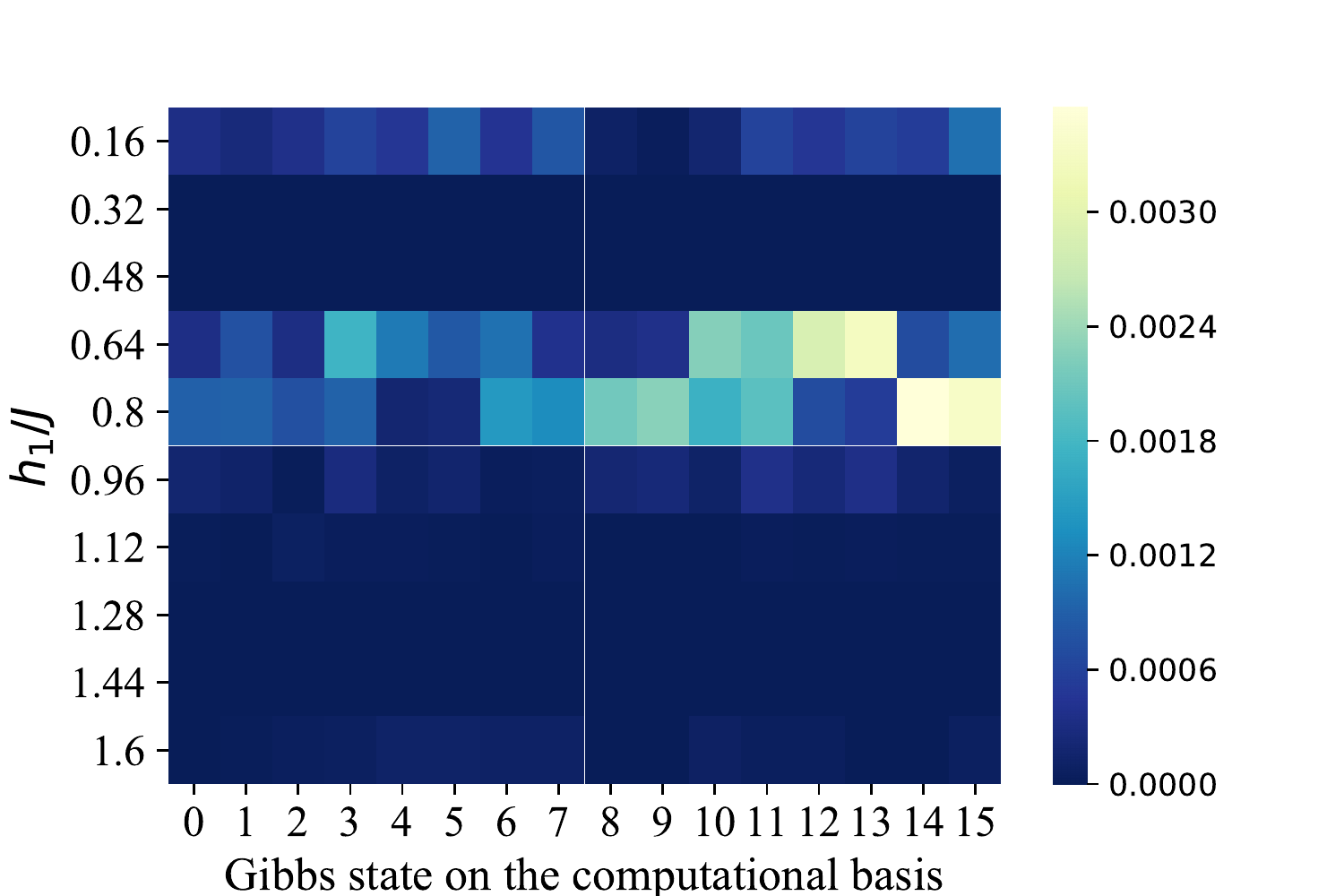}
  }
  \subfigure[(b)]{
  \includegraphics[width=0.48\textwidth]{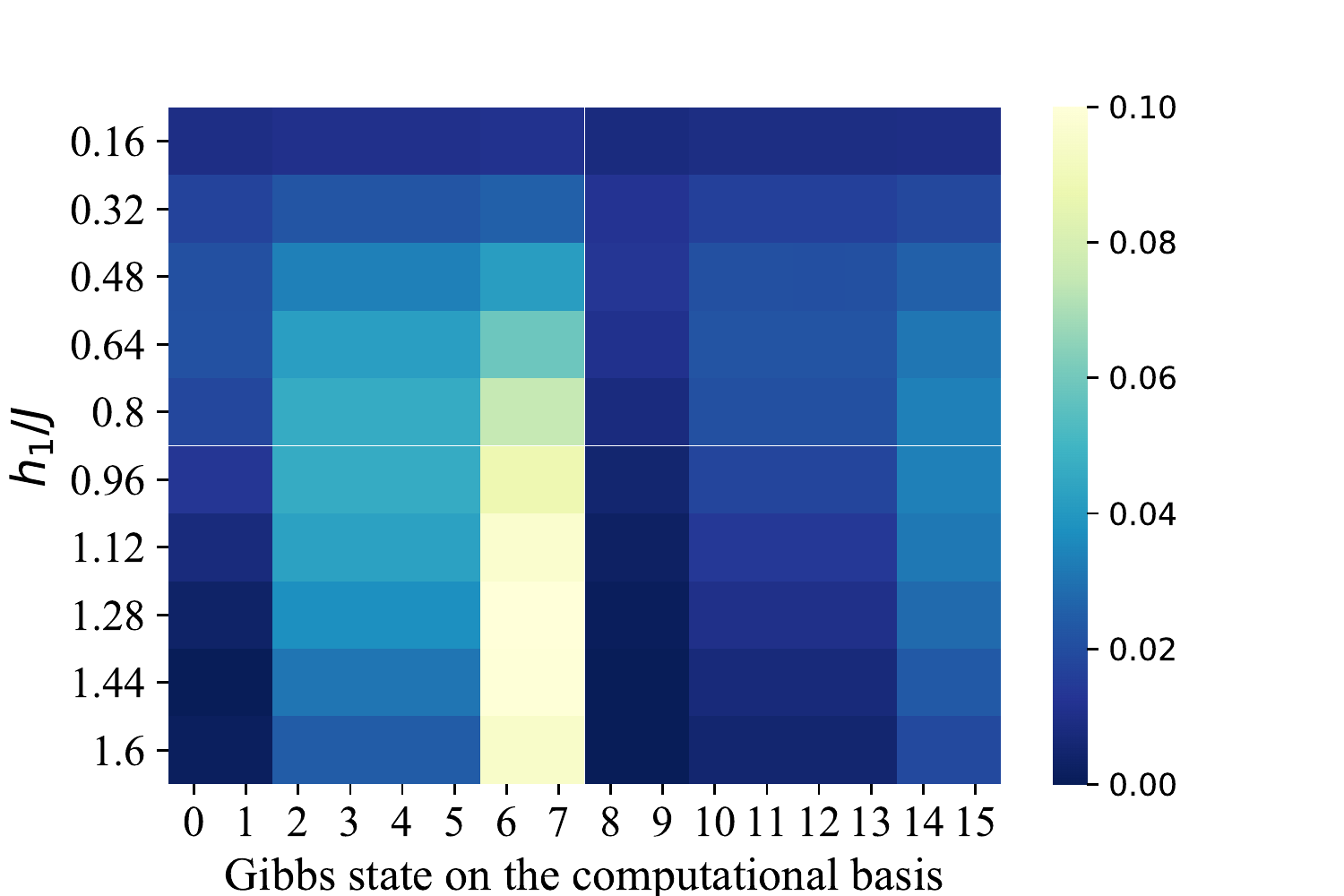}
  }
  \DeclareGraphicsExtensions.
  \caption{The Gibbs states (thermal states) of the Haldane chain in the case of $N=9, h_2=0, h_1/J\in[0.16,1.60]$, which are computed by the (a) 2L-QRBM and (b) the hardware-efficient ansatz, respectively. The darker pixels indicate smaller error with the exact value, while the lighter pixels show the opposite. We sample 16 computational basis from the total 512 basis by implementing a large number of measurements (100,000). The simulation error of 2L-QRBM achieves nearly $\epsilon=\mathcal{O}(10^{-4})$.}
  \label{Fig:GibbsState}
\end{figure}

\noindent \textbf{Prepare the 2L-QRBM by quantum circuit.} We now show how to design a quantum circuit for preparing the 2L-QRBM trial state with the Quantum Imaginary Time Evolution (QITE) algorithm \cite{Lehto2007QITE}. Noting that the Hamiltonian $\mathcal{H}_{RBM}(\bm{\theta})=\sum_{s}\widehat{h}_s(\bm{\theta})$ is composed by the linear combination of  operators that act on at most $k$ qubits $(k=1,2)$. According to the Trotter theorem \cite{Nielsen2002Quantum}, the operator $e^{\mathcal{H}_{RBM}(\bm{\theta})}$ can be decomposed as:
\begin{align}
e^{\mathcal{H}_{RBM}(\bm{\theta})}=(e^{\widehat{h}_1(\bm{\theta})/n}e^{\widehat{h}_2(\bm{\theta})/n}...)^n+\mathcal{O}\left(\frac{1}{n^2}\right),
\end{align}
in which the parameter $n$ is the number of Trotter steps. For the $s$-term $\widehat{h}_s(\bm{\theta})$, after a single Trotter step, the initial state $|\Psi_0\rangle$ becomes to
\begin{align}
|\Psi\rangle=c^{-1/2}e^{\widehat{h}_s(\bm{\theta})/n}|\Psi_0\rangle,
\label{Eq:SingleTrotter}
\end{align}
and the normalization parameter $c$ can be estimated by $c=1-\frac{2}{n}\langle\Psi_0|\widehat{h}_s(\bm{\theta})|\Psi_0\rangle+\mathcal{O}(1/n^2)$ according to the truncated Taylor series. To implement Eq.(\ref{Eq:SingleTrotter}) on the NISQ devices, Chan et al. \cite{Mario2019QITE} introduces a unitary operator $e^{-iA_s(\bm{\theta})/n}$ to approximate it, where the operator $A_s(\bm{\theta})$ (acting on $k$ qubits) can be extended in the Pauli basis with relevant parameters $a_s(\bm{\theta})_{i_1...i_k}$
\begin{align}
A_s(\bm{\theta})=\sum\limits_{i_1...i_k}a_s(\bm{\theta})_{i_1...i_k}\sigma_{i_1}...\sigma_{i_k}.
\end{align}
Define $|\Delta_0\rangle=n(|\Psi\rangle-|\Psi_0\rangle)$ and $|\Delta\rangle=-iA_s(\bm{\theta})|\Psi_0\rangle$, the goal is to find out an optimal operator $A_s(\bm{\theta})$ to
minimize $\||\Delta_0\rangle-|\Delta\rangle\|$. Taking parameters $a_s(\bm{\theta})_{i_1...i_k}$ as variables, finding out $A_s(\bm{\theta})$ can be recognized as an optimization procedure, and parameters $a_s(\bm{\theta})_{i_1...i_k}$ can be efficiently determined by solving the linear equation $(\bm{S}+\bm{S}^{\dagger})\bm{a_s(\bm{\theta})}=-\bm{b}$, where the matrix entries $S_{i_1...i_k,j_1...j_k}=\langle\Psi_0|\sigma_{i_1...i_k}^{\dagger}\sigma_{j_1...j_k}|\Psi_0\rangle$ and vector entries $b_{i_1...i_k}=-ic^{-1/2}\langle\Psi_0|\sigma_{i_1...i_k}^{\dagger}$ $\widehat{h}_s(\bm{\theta})|\Psi_0\rangle$. All of the entries $S_{i_1...i_k,j_1,...,j_k}$ and $b_{i_1...i_k}$ can be efficiently estimated by the swap test method by the implementing of $\mathcal{O}(1/\epsilon^2)$ measurements with an acceptable error $\epsilon$. This optimization problem can be efficiently solved by a classical computer once all entries $S_{i_1...i_k,j_1,...,j_k}$ and $b_{i_1...i_k}$ are all estimated by a quantum computer (though the computational overhead of solving the linear equation is $\mathcal{O}(\textrm{poly}(2^k))$), it does not serve as the dominant component because $k$ takes value from $\{1,2\}$. Therefore, the complexity of implementing the 2L-QRBM by using QITE algorithm is quasi-polynomial in $n$ (the number of Trotter steps) \cite{Mario2019QITE}.

\noindent\textbf{Experimental results.} To solve different tasks, the 2L-QRBM should be trained by different constraints. Given a Hamiltonian $\mathcal{H}=\sum_{j}\alpha_jP_j$, where $\alpha_j\in\mathcal{R}$ and $P_j=P_j^1\otimes P_j^2\otimes...\otimes P_j^N, P_j^s\in\{I,X,Y,Z\}$, we here propose two training methods for 2L-QRBM to compute the ground state energy and Gibbs state of $\mathcal{H}$, respectively.

\emph{Compute ground state energy.} We first give the elaborate details for training the ansatz to compute the ground state energy. Similar to the variational quantum eigen-solver (VQE) algorithm, the ansatz $|\Psi_v(\bm{\theta})\rangle$ approximates the ground state along with the energy $E(\bm{\theta})=\langle\Psi_v(\bm{\theta})|\mathcal{H}|\Psi_v(\bm{\theta})\rangle$ is minimized (and meanwhile the Boltzmann parameters theta is trained/updated) by the iterative optimization method. We here utilize a kind of gradient-descent method called ``the Simultaneous Perturbation Stochastic Approximation (SPSA) algorithm'' to optimize the cost function. SPSA algorithm is robust against the statistical fluctuations, and has shown the merits of high accuracy in the optimization of cost function \cite{Kandala2018QVE}. Concretely, in every step (e.g. the $k$-th step) of the SPSA algorithm, the gradient at $\bm{\theta}_k$ is constructed as $g_k(\bm{\theta}_k)=(E(\bm{\theta}_k^+)-E(\bm{\theta}_k^-))\Delta_k/2c_k$, where $\bm{\theta}_k^{\pm}=\bm{\theta}_k\pm c_k\Delta_k$, $\Delta_k$ is sampled according to the Bernoulli distribution and $c_k$ can be selected with priori experience.


We here simulate the ground state energy of the Hydrogen molecular as well as the Water molecular on ProjectQ, and compares the performance of our 2L-QRBM and previous works including hardware-efficient ansatz \cite{Kandala2018QVE}, QRBM without the transverse field and the Full
Configuration Interaction (FCI) method. The numerical results (see Fig.\ref{Fig:GroundEnergy}) shows that, the QRBM without the transverse field and the hardware-efficient ansatz cannot converge to an optimal solution when the bond length increases from $1.5$ to $2.5$ (both in (a) and (b)), and the curves of our 2L-QRBM are extremely close to that of FCI method.
\begin{figure}
  \begin{center}
  \includegraphics[width=0.9\textwidth]{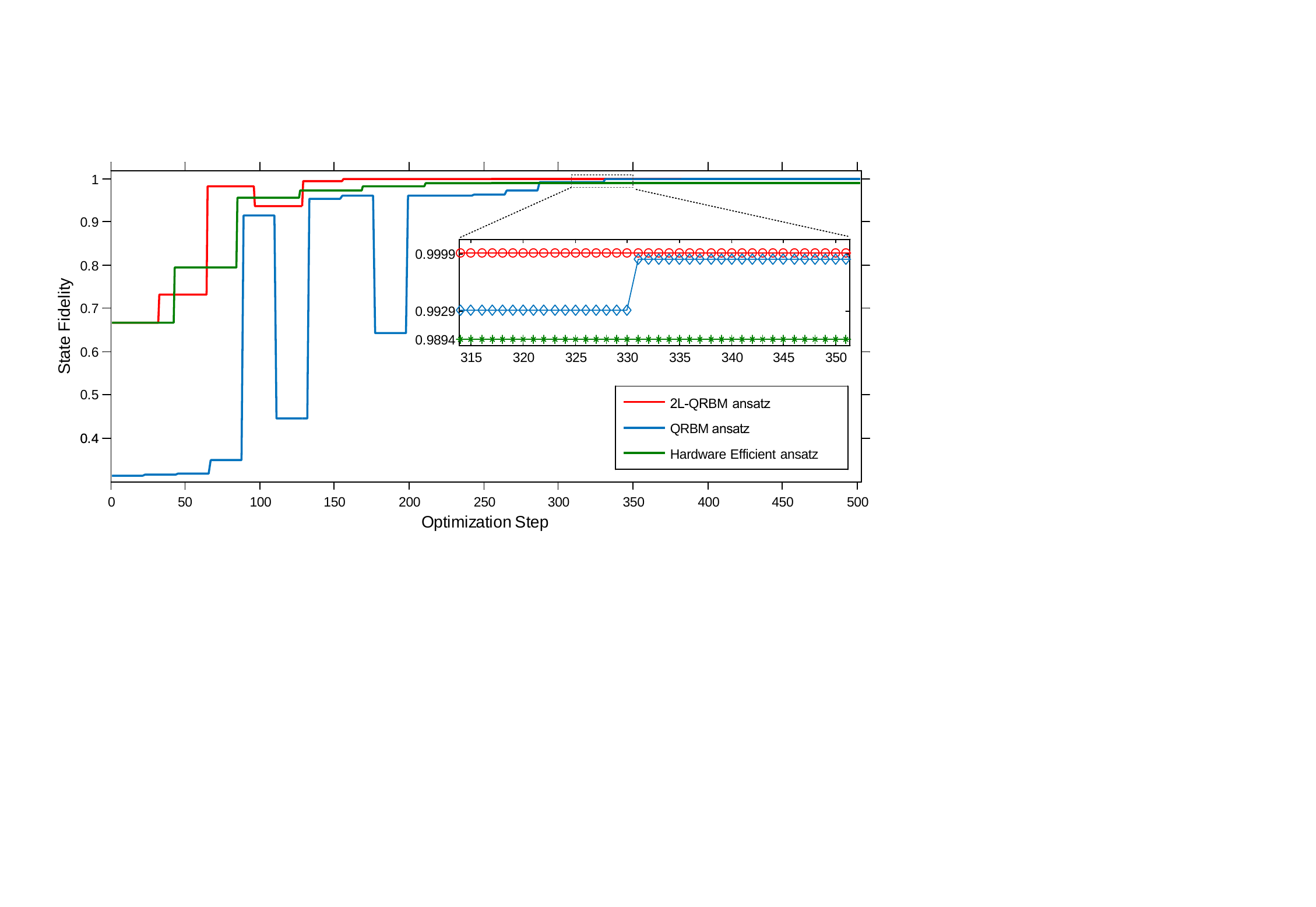}\\
  \caption{Finding the Gibbs state of Haldane chain with parameters $N=9, h_1/J=0.48$ and $h_2=0$ by using 2L-QRBM (red line) versus QRBM (blue line) and hardware-efficient ansatz (green line). In this case,
  the case of  2L-QRBM converges more rapidly to the target state (nearly with the fidelity of $100\%$) than the other two methods.
  }\label{GibbsCurve}
  \end{center}
\end{figure}
\begin{figure}

  \begin{center}

  \includegraphics[width=0.6\textwidth]{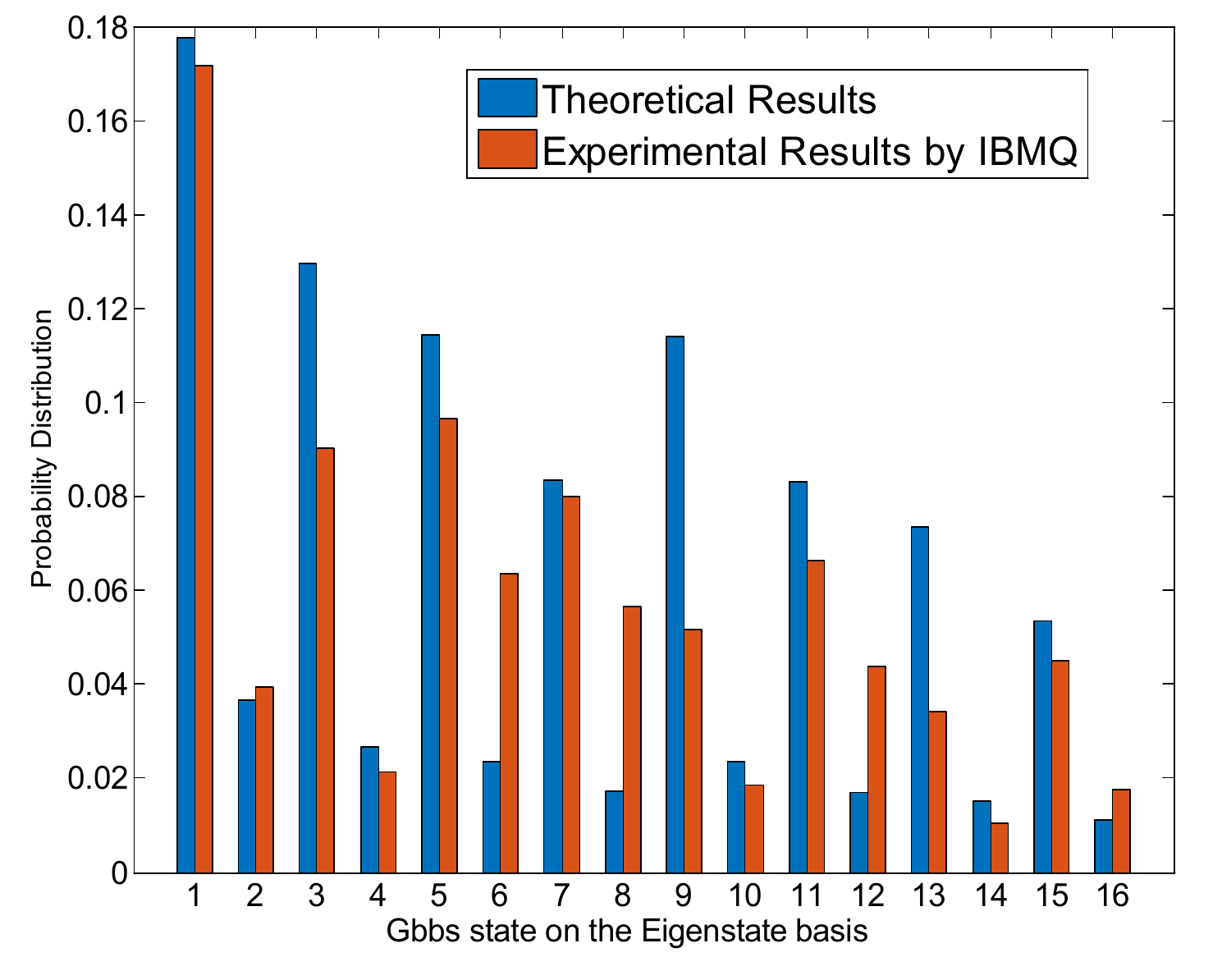}\\
  \caption{The experimental results by using 2L-QRBM to compute the Gibbs state of Haldane chain on the ibmq-essex chip when $N=4, h_2=0$ and $h_1=0.48$. The frequency of the utilized four superconductor qubits are ranging from $[4.4997(GHz), 4.6946 (GHz)]$, the maximum error rate of single-qubit-gate is $4.56\times10^{-4}$, and that for C-NOT gate is $1.474\times10^{-2}$ \cite{Qskit2019}.}\label{GibbsBar}
  \end{center}
\end{figure}

\emph{Compute Gibbs state.} We now propose the training method to compute the Gibbs state (thermal state) $\rho=e^{-\beta \mathcal{H}}/\mathcal{Z}$ of the Hamiltonian $\mathcal{H}=\sum_{j}\alpha_jP_j$. In general, Gibbs state comes from the process that performing $e^{-\beta \mathcal{H}/2}$ onto the first system of the maximally mixed state $|\phi\rangle=2^{-N/2}\sum_x|x\rangle|x\rangle$, followed by tracing out the second system. We here aim to tune the trial state $|\Psi_v(\bm{\theta})\rangle$ to approximate the state $|\phi_\tau\rangle=c^{-1/2}e^{-\beta \mathcal{H}/2}|\phi_0\rangle$, where $\tau=1/\beta$. There might be a doubt why we do not invoke the QITE algorithm to implement $|\phi_\tau\rangle$ directly. Actually, if the target Hamiltonian is $k$-local ($k$ is not too large), QITE algorithm can efficiently solve the quantum Gibbs sampling problem, otherwise it will induce an enormous computation overhead because it needs to solve  a linear function $(\bm{S}+\bm{S}^{\dagger})\bm{a_s(\bm{\theta})}=-\bm{b}$. Fortunately, our 2L-QRBM naturally provides a 2-local Hamiltonian whose ground state can approximate the purified Gibbs state of a k-local Hamiltonian, even when $k$ is large.


We introduce the Wick rotation $(t\rightarrow i\tau)$ and the VQE algorithm to adjust the parameter so that the trial state $|\Psi_v(\bm{\theta})\rangle$ gradually approximates the purified Gibbs state $|\phi_\tau\rangle$. The optimal Boltzmann parameter is obtained when the equation $\delta\|(\partial/\partial\tau+\mathcal{H}-E_{\tau})|\Psi_v(\bm{\theta})\rangle\|=0$ is satisfied, where $E_{\tau}$ represents the energy term. In this case, the parameter $\bm{\theta}$ can be obtained from the equation $\theta(\tau+\delta\tau)=\theta(\tau)+A^{-1}(\tau)C(\tau)\delta\tau$, where the elements of matrix $A$ are defined as $A_{mn}=\Re\langle\partial\Psi_v(\bm{\theta})/\partial\theta_n|\partial\Psi_v(\bm{\theta})/\partial\theta_m\rangle$, that of $C$ are $C_n=-\Re\langle\partial_{\theta_n}\Psi_v(\bm{\theta})|\mathcal{H}|\Psi_v(\bm{\theta})\rangle$ \cite{MacArdle2019VarITE}. the notation $\Re(x)$ represents the real part of $x$.

Following the procedure above, we compute the Gibbs state of a family of Hamiltonians on a spin $-1/2$ chain with the open boundary conditions
\begin{align}
\mathcal{H}=-J\sum\limits_{i=1}^{N-2}Z_iX_{i+1}Z_{i+2}-h_1\sum\limits_{i=1}^NX_i-h_2\sum\limits_{i=1}^{N-1}X_iX_{i+1},
\end{align}
where $h_1, h_2$ and $J$ are the changeable parameters of the Hamiltonian. We compute the cases that $N=9, h_2=0, h_1/J\in[0.16,1.60]$ by using the 2L-QRBM ansatz and the hardware-efficient ansatz with swap entanglement operator on the ProjectQ, respectively. We implement 100,000 measurements on the computational basis to reconstruct the target Gibbs state. The results are illustrated as Fig.\ref{Fig:GibbsState} (a) (2L-QRBM) and (b) (hardware-efficient). Each row of the graph represents the amplitude differences $||P^{trial}_i-P^{exact}_i||$ between the trial state and the target one on the 16 sampled computational basses (from 512 possible basis), where $i\in\{0,1,...,2^N\}$. Obviously, the error rate induced in 2L-QRBM is nearly $\mathcal{O}(10^{-4})$, which is much less than that of the hardware-efficient ansatz. We also illustrate the optimization steps when computing the Gibbs state of Haldane chain for the specific parameters $N=9, h_1/J=0.48$ and $h_2=0$ (see Fig.\ref{GibbsCurve}). It is evident that 2L-QRBM converges to an appropriate destination faster than the other two methods.

Finally, we test the 2L-QRBM model on the IBMQ-essex quantum device by computing the Gibbs state of the Haldane chain (see Fig.\ref{GibbsBar}). Limited by the single- and double- gate error rates, the experimental fidelity is only $0.926$ which is below the corresponding theoretical value. We believe that, with the rapidly refinement of the quantum hardware devices, our 2L-QRBM model can simulate much more functions with high accuracy.
\section*{Discussion}
In summary, we propose a QRBM induced by a specially designed 2-local Hamiltonian, and our model has two salient features, as stated above. First, it is proved that the 2L-QRBM can implement the universal quantum computation tasks, meanwhile our proof implies a new way for understanding the QRBM. Actually, 2L-QRBM can be recognized as the variational version of the QITE algorithm, and this relationship is similar to that for the Quantum Approximation Optimization Algorithm (QAOA) and the Adiabatic Quantum Computation (AQC). We hope our proof can inspire more interesting perspectives on understanding quantum machine learning algorithms in terms of the representational power. Second, our model can be efficiently transformed to a quantum circuit and be implemented on the NISQ devices. Different from the AQC whose computational complexity depends on the depth of the corresponding quantum circuit, the complexity for implementing the 2L-QRBM only comes from the Trotter steps. The exact simulation results show the 2L-QRBM trial state can converge to a better solution compared with the widely utilized hardware-efficient ansatz and the Ising model based QRBM. We also test our model on the superconducting quantum devices with $4$ qubits, and the experimental results illustrate 2L-QRBM can approximate the target wave function with an acceptable error on the realistic devices. One of the possible reasons for high fidelity performance is that the 2L-QRBM is sufficient for universal quantum computation tasks, therefore the target quantum state is bounded by the Hilbert space that 2L-QRBM can depict. This superior performance shows our model not only has theoretical significance, but also has application values in the NISQ era.

\section*{Methods}
\textbf{Review of the RBM and QRBM.} As a classic machine learning technique, the RBM serves as the basis of complex deep learning models such as deep belief networks and deep Boltzmann machines \cite{Hinton2006DBN}. It comprises a probabilistic network of binary units with a quadratic energy function. The RBM are commonly constituted by one visible layer of $N$ nodes $\mathbf{v}=\{v_i\}_{i=1}^N$, corresponding to the physical spin variables in a chosen basis and a single hidden layer of $M$ auxiliary nodes $\mathbf{h}=\{h_j\}_{j=1}^M$. To maintain consistency with the standard notation in quantum mechanics, the units $v_i$ and $h_j$ take value from $\{0,1\}$, and the corresponding energy function is a linear combination of them, that is,
\begin{align}
\label{Eq:Energy}
E_{\bm{\theta}}(\textbf{v,h})=\sum\limits_{i=1}^Nb_iv_i+\sum\limits_{j=1}^Mm_jh_j+\sum\limits_{i=1}^N\sum\limits_{j=1}^MW_{ij}v_ih_j,
\end{align}
where $\bm{\theta}=\{b_i,m_j,W_{ij}\}$ are Boltzmann parameters. The (unnormalized) marginal distribution of observing a visible variable $\mathbf{v}$ is given by $\Psi_{\mathbf{v}}(\bm{\theta})=\sum_{\mathbf{h}}e^{-E_{\bm{\theta}}(\textbf{v,h})}$. Utilizing the RBM to fit a target wave function can be achieved by the minimization of the loss function $\mathcal{L}(\bm{\theta})$ via the tuning of Boltzmann parameters $\bm{\theta}$, and the form of $\mathcal{L}(\bm{\theta})$ depends on the realistic problem.

In the problem of computing the ground state of a Hamiltonian $\mathcal{H}$, the loss function can be chosen as $\mathcal{L}(\bm{\theta})=\langle\Psi(\bm{\theta})|\mathcal{H}|\Psi(\bm{\theta})\rangle/\langle\Psi(\bm{\theta})|\Psi(\bm{\theta})\rangle$. Here, $|\Psi(\bm{\theta})\rangle=\sum_{\mathbf{v}}\Psi_{\mathbf{v}}(\bm{\theta})|\mathbf{v}\rangle$ is a superposition state, corresponding to the $2^N$ possible inputs of $\mathbf{v}$. The RBM parameter $\bm{\theta}$ as well as the amplitudes $\Psi_{\mathbf{v}}(\bm{\theta})$ are tuned in the minimization of $\mathcal{L}(\bm{\theta})$ so that $|\Psi(\bm{\theta})\rangle$ converges to the ground state of $\mathcal{H}$.

To enhance the computational efficiency of the RBM, the QRBM model are proposed \cite{RongQMLESC2018,Shengyu2019UCRBM,Carleo2020QRBM}. Different from the RBM, the QRBM utilizes a quantum circuit to parallelly compute the amplitudes $\Psi_{\mathbf{v}}(\bm{\theta})$, and naturally outputs the superposition state $|\Psi(\bm{\theta})\rangle=\sum_{\mathbf{v}}\frac{\Psi_{\mathbf{v}}(\bm{\theta})}{\mathcal{Z}}|\mathbf{v}\rangle$,
where $\mathcal{Z}=\sum_{v}\Psi_{\mathbf{v}}^2(\bm{\theta})$ is the normalized factor.

\noindent \textbf{Self energy and effective Hamiltonian.}
As we concerns only the low energy and ground state of the Hamiltonian, we now illustrate some methods to approximate the Hamiltonian in the low energy space (details refer to \cite{Kitaev2005CAQC}).

Given a Hamiltonian $\mathcal{H}$, the Hilbert space $\mathcal{H}^{Space}$ can be divided as $\mathcal{H}^{Space}=\mathcal{L}_+\oplus\mathcal{L}_-$, where $\mathcal{L}_+$ is the space spanned by the eigenvectors of $\mathcal{H}$ with eigenvalues $\lambda\geq\lambda_c$ and $\mathcal{L}_-$ is spanned by that with $\lambda<\lambda_c$. Let $\Pi_{\pm}$ be the corresponding projection operators onto $\mathcal{L}_{\pm}$. Given an operator $X$ on the Hilbert space $\mathcal{H}^{Space}$, $X_{++}$ is defined as $\Pi_+X\Pi_+$, which is an operator on $\mathcal{L_+}$, and similarly $X_{--}=\Pi_-X\Pi_-$ is an operator on the low energy subspace $\mathcal{L}_-$.

The self energy of $\mathcal{H}$ is defined as
\begin{align}
\Sigma_{-}(z)=zI_{-}-G^{-1}_{--}(z),
\end{align}
where $G(z)=(zI-\mathcal{H})^{-1}$. $G(z)$ is a meromorphic operator-valued function of the complex variable $z$ with poles at $z=\lambda_j$, where $\lambda_j$ is the eigenvalue of $\mathcal{H}$.
The self energy $\Sigma_{-}(z)$ is utilized to approximate the spectrum of $\mathcal{H}$ in the low energy subspace. Note that $\Sigma_{-}(z)$ is nearly constant for a certain range of $z$, therefore a Hamiltonian $\mathcal{H}_{eff}$ can be selected to approximate it. That is, $\mathcal{H}_{eff}$ can approximate $\mathcal{H}$ in the low energy space, so it is generally called ``effective Hamiltonian" in the computation of $\mathcal{H}$'s ground state and ground energy.

We now give several theorems to establish the relationship between our 2L-QRBM and quantum circuit.

\textbf{Theorem 2.} \emph{Given an arbitrary quantum circuit on $n$ qubits with $l$ layers of single-qubit or two-qubit gates implementing a unitary $U$, suppose $|\alpha(l)\rangle$ is the output of this circuit, then for an arbitrarily $\epsilon>0$, there exists a $2$-local Hamiltonian $\mathcal{H}$ whose ground state $|\psi\rangle$ is $\mathcal{O}((4/\Delta_{eff}+1/\sqrt{L})\epsilon)$ close to the state $|\alpha(l)\rangle$, where $\Delta_{eff}$ is the spectral gap of $\mathcal{H}$'s effective Hamiltonian $\mathcal{H}_{eff}$.}

We first review several lemmas in \cite{Aharonov2005AQC, Kitaev2005CAQC}, which are closely related with our proof of Theorem 2.

\emph{\textbf{Lemma 2.1}}\cite{Aharonov2005AQC}. \emph{Given a quantum circuit on $n$ qubits with $L$ two-qubit gates implementing a unitary $U$, and $\epsilon>0$, there exists a 3-local Hamiltonian $\mathcal{H}_{final}$ whose ground state is $\mathcal{O}(\epsilon/\sqrt{L})$ close (in trace distance) to the quantum state $U|0\rangle^{n}$. Moreover, the Hamiltonian $\mathcal{H}_{final}$ can be computed by a polynomial time Turing machine.}

\emph{\textbf{Lemma 2.2}}\cite{Kitaev2005CAQC}. \emph{Suppose $\mathcal{H}$ is a Hamiltonian with a spectral gap $\Delta$ around the cutoff  $\lambda_c$ (that is, all its eigenvalues are in $(-\infty,\lambda_-]\cup[\lambda_+,+\infty)$, where $\lambda_+=\lambda_c+\Delta/2$ and $\lambda_-=\lambda_c-\Delta/2$), and $V$ is a Hermite operator with norm $\|V\|\leq\Delta/2$, then for an arbitrarily small positive value $\epsilon$, if there exists an operator $\mathcal{H}_{eff}$ whose eigenvalues belongs to $[c,d]$ for some $c<d<\lambda_c-\epsilon$ and moreover, the inequality
\begin{align}
\|\Sigma_-(z)-\mathcal{H}_{eff} \|\leq\epsilon
\end{align}
(where $\Sigma_-(z)$ is the self energy of $\widetilde{\mathcal{H}}=\mathcal{H}+V$) holds for all $z\in[c-\epsilon,d+\epsilon]$, each eigenvalue $\lambda_j$ is $\epsilon$ close to the $j$-th eigenvalue of $\mathcal{H}_{eff}$.}
\emph{\textbf{Lemma 2.3}}\cite{Kitaev2005CAQC}. \emph{Assume that $\mathcal{H}, V, \mathcal{H}_{eff}$ satisfy the conditions of Lemma 2 with some $\epsilon_2>0$, let $\lambda_{eff,i}$ denote the $i$-th eigenvalue of $\mathcal{H}_{eff}$ and $|\widetilde{v}\rangle$ (resp., $|v_{eff}\rangle$) denote the ground state of $\widetilde{\mathcal{H}}(resp., \mathcal{H}_{eff})$. Suppose $\lambda_{eff,2}$ and $\lambda_{eff,1}$ are the smallest two eigenvalues of $\mathcal{H}_{eff}$ and $\lambda_{eff,2}>\lambda_{eff,1}$, then we have}
\begin{align}
\||\widetilde{v}\rangle-|v_{eff}\rangle\|\leq\frac{2\|V\|^2}{(\lambda_+-\lambda_{eff,1}-\epsilon_2)^2}+\frac{4\epsilon_2}{\lambda_{eff,2}-\lambda_{eff,1}}.
\end{align}

We begin our proof of Theorem 2 now.  Given an arbitrary $l$-layer quantum circuit, without loss of generality, we suppose the input state of this circuit is $|0\rangle^{\otimes n}$ and the output state is
$|\alpha(l)\rangle$. According to Lemma 2.1, for an arbitrary small positive value $\epsilon_1$, there exists a $3$-local Hamiltonian $\mathcal{H}^{(3)}$ whose ground state $|v^{(3)}\rangle$ is $\mathcal{O}(\epsilon_1/\sqrt{L})$ close to $|\alpha(l)\rangle$, that is,
\begin{align}
\||v^{(3)}\rangle-|\alpha(l)\rangle\|=\epsilon_1/\sqrt{L}.
\label{Eq:3localHistoryState}
\end{align}
It is interesting to note that any $3$-local Hamiltonian $\mathcal{H}^{(3)}$ can be represented as \cite{Kitaev2005CAQC}
\begin{align}
\mathcal{H}^{(3)}=Y-6\sum\limits_{m=1}^MB_{m1}B_{m2}B_{m3},
\label{Eq:H3}
\end{align}
where $Y$ is a 2-local Hamiltonian with the norm bound $\mathcal{O}(1/n^6)$, $M=\mathcal{O}(n^3)$, $n$ is the scale of the quantum system, and each $B_{mi}\geq \frac{1}{n^3}I$ is a linear combination of the Pauli operators. Now we construct a 2-local Hamiltonian $\mathcal{H}^{(2)}$ whose ground state can approximate that of $\mathcal{H}^{(3)}$. For an arbitrarily small positive value $\delta$, let
\begin{align}
\mathcal{H}=-\frac{\delta^{-3}}{4}\sum\limits_{m=1}^MI\otimes(\sigma^z_{m1}\sigma^z_{m2}+\sigma^z_{m1}\sigma^z_{m3}+\sigma^z_{m2}\sigma^z_{m3}-3I),
\end{align}
and
\begin{align}
V=Y\otimes I+\delta^{-1}\sum\limits_{m=1}^M(B^2_{m1}+B^2_{m2}+B^2_{m3})\otimes I-\delta^{-2}\sum\limits_{m=1}^M(B_{m1}\otimes\sigma^x_{m1}+B_{m2}\otimes\sigma^{x}_{m2}+B_{m3}\otimes\sigma_{m3}^x),
\label{Eq:V}
\end{align}
then $\mathcal{H}^{(2)}=\mathcal{H}+V$ is a 2-local Hamiltonian, and its self energy can be written as
\begin{align}
\Sigma_-(z)=Y\otimes I-6\sum\limits_{m=1}^MB_{m1}B_{m2}B_{m3}\otimes(\sigma^x)_{eff}+\mathcal{O}(\delta).
\end{align}
Let $\mathcal{H}_{eff}=Y\otimes I-6\sum_{m=1}^MB_{m1}B_{m2}B_{m3}\otimes(\sigma^x)_{eff}$, the self energy of $\mathcal{H}^{(2)}$ can be rewritten as $\Sigma_-(z)=\mathcal{H}_{eff}+\mathcal{O}(\delta)$.
Since $\|\mathcal{H}_{eff}\|\leq\mathcal{O}(1)$ and $\|V\|=\mathcal{O}(\delta^{-2})$, applying Lemma 2.2 with $c=-\|\mathcal{H}_{eff}\|$, $d=\|\mathcal{H}_{eff}\|$, and $\lambda_c=\Delta/2$, where $\Delta=\delta^{-3}$ is the spectral gap of $\mathcal{H}$, we can obtain that the smallest eigenvalue of $\mathcal{H}^{(2)}$ is $\mathcal{O}(\delta)$ close to that of $\mathcal{H}_{eff}$.

We now exploit the relationship between $\mathcal{H}_{eff}$ and $\mathcal{H}^{(3)}$. Note that
\begin{align}
\mathcal{H}_{eff}&=\left(Y-6\sum\limits_{m=1}^MB_{m1}B_{m2}B_{m3}\right)\otimes|+\rangle\langle+|+\left(Y+6\sum\limits_{m=1}^MB_{m1}B_{m2}B_{m3}\right)\otimes|-\rangle\langle-|\\
&=\mathcal{H}^{(3)}\otimes|+\rangle\langle+|+\left(Y+6\sum\limits_{m=1}^MB_{m1}B_{m2}B_{m3}\right)\otimes|-\rangle\langle-|
\label{Eq:Effective}
\end{align}
and $B_{m1}B_{m2}B_{m3}\geq0$, then the ground state of $\mathcal{H}_{eff}$ can be written as $|v_{eff}\rangle=|v^{(3)}\rangle|+\rangle$, where $|v^{(3)}\rangle$ is the ground state of $\mathcal{H}^{(3)}$.

Note that the 2-local Hamiltonian $\mathcal{H}^{(2)}$, Hermite operator $V$ (Eq.\ref{Eq:V}) and effective Hamiltonian $\mathcal{H}_{eff}$ (Eq.\ref{Eq:Effective}) satisfy the conditions in Lemma 2.2, then according to Lemma 2.3, we have
\begin{align}
\||v^{(2)}\rangle-|v_{eff}\rangle\|\leq\frac{2\|V\|^2}{(\lambda_+-\lambda_{eff,1}-\epsilon_2)^2}+\frac{4\epsilon_2}{\lambda_{eff,2}-\lambda_{eff,1}},
\end{align}
where $|v^{(2)}\rangle$ is the ground state of $\mathcal{H}^{(2)}$. Since $|v_{eff}\rangle=|v^{(3)}\rangle|+\rangle$ and $\||v^{(3)}\rangle-|\alpha(l)\rangle\|=\epsilon_1/\sqrt{L}$ (Eq.\ref{Eq:3localHistoryState}), we have
\begin{align}
\||v^{(2)}\rangle-|\alpha(l)\rangle|+\rangle\|&\leq\||v^{(2)}\rangle-|v_{eff}\rangle\|+\||v_{eff}\rangle-|\alpha(l)\rangle|+\rangle\|\\
&=\||v^{(2)}\rangle-|v_{eff}\rangle\|+\||v^{(3)}\rangle|+\rangle-|\alpha(l)\rangle|+\rangle\|
\end{align}
\begin{align}
\leq\frac{2\|V\|^2}{(\lambda_+-\lambda_{eff,1}-\epsilon_2)^2}+\frac{4\epsilon_2}{\lambda_{eff,2}-\lambda_{eff,1}}+\frac{\epsilon_1}{\sqrt{L}}
=\epsilon(4/\Delta_{eff}+1/\sqrt{L})=\mathcal{O}(\epsilon),
\end{align}
where $\Delta_{eff}=\lambda_{eff,2}-\lambda_{eff,1}$, $\epsilon=\max\{\epsilon_1,\epsilon_2\}=\mathcal{O}(\delta)$, $||V||=\mathcal{O}(\delta^{-2})$, $\lambda_+=\delta^{-3}$, and $|\lambda_{eff,1}|=\mathcal{O}(1)$.

\textbf{Theorem 3.} \emph{For an arbitrary 2-local Hamiltonian $\mathcal{H}$, there exists a simplified 2-local Hamiltonian in the form of
\begin{align}
\mathcal{H}_{s}\bm{(\widetilde{\theta})}=\sum_{i=1}^{N}\sum_{t\in\{x,y,z\}}b_i^tv_i^t+\sum_{s=1}^{N-1}\sum_{k=s+1}^N\sum_{t\in\{x,y,z\}}K^{t}_{sk}v_s^tv_k^t
\label{Eq:simplified H}
\end{align}
which can approximate $\mathcal{H}$ in the low energy subspace, that is, the Hamiltonian only with interaction terms $v_i^s\otimes v_j^l$, $s\neq l\in\{x,y,z\}$ can be approximated by a Hamiltonian with terms $\{v_i^s, v_i^s\otimes v_j^s\}$ in the low energy subspace, where $s\in\{x,y,z\}$.
}

\noindent \textbf{Proof.}
Jocob et al.\cite{Jacob2008} showed that the intersection $\sigma_i^z\sigma_j^x$ can be constructed from $\sigma^x\sigma^x$ and $\sigma^z\sigma^z$ in the low energy subspace. Following their method, we here propose how to approximate terms $\sigma_i^x\sigma_j^y$ and $\sigma_i^z\sigma_j^y$ by merely using terms $\{\sigma_i^s, \sigma_i^s\otimes\sigma_j^s\}_{i,j=1}^N$, where $s\in\{x,y,z\}$.


Given a 2-local Hamiltonian $\alpha_{ij}\sigma_i^x\sigma_j^y$ and an arbitrary $\delta>0$, our target is to find a 2-local Hamiltonian, composed only by $\sigma^x, \sigma^y, \sigma^x\sigma^x$ and $\sigma^y\sigma^y$, which can be $\mathcal{O}(\delta)$ close to  $\alpha_{ij}\sigma_i^x\sigma_j^y$ in the low energy space. Let
\begin{align}
\mathcal{H}^{(2)}=\sigma_i^x+\sigma_j^x+\sigma_i^y+\sigma_j^y+\sigma_i^x\sigma_j^x+\sigma_i^y\sigma_j^y,
\end{align}
\begin{align}
V_1=(\mathcal{H}^{(2)}+D(\sigma^y_j+I))\otimes I_k-A\sigma_i^x\otimes|-\rangle\langle-|_k,
\end{align}
\begin{align}
V_2=B(\sigma_j^y\otimes I)\otimes\sigma_k^y,
\end{align}
\begin{align}
V_3=C\sigma_i^x\otimes|+\rangle\langle+|_k,
\end{align}
and $V=V_1+V_2+V_3$, where $A,B,C,D$ are real-valued parameters to be determined later. Then, the self energy operator $\Sigma_-(z)$ of the 2-local Hamiltonian $\mathcal{H}^{(2)}+V$ can be written as
\begin{eqnarray}
\begin{split}
\Sigma_{-}(z)=\widetilde{\mathcal{H}}^{(2)}+\left(\frac{2B^2C}{(z-\delta^{-1})^2}-A\right)\sigma_i^x+\left(\frac{2B^2}{z-\delta^{-1}}+D+\frac{4DB^2}{(z-\delta^{-1})^2}\right)(\sigma_j^y+I)+\frac{2B^2C}{(z-\delta^{-1})^2}\sigma_i^x\sigma_j^y+\mathcal{O}(\delta^{3}),
\end{split}
\end{eqnarray}
where $\widetilde{\mathcal{H}}^{(2)}=\mathcal{H}^{(2)}+\frac{B^2}{(z-\delta^{-1})^2}(\sigma_j^y+I)\mathcal{H}^{(2)}(\sigma_j^y+I)$. Select a random nonzero real number $E$, and let
\begin{align}
h_i=\left(1+\frac{2B^2}{(z-\delta^{-1})^2}\right),
\end{align}
\begin{align}
\Delta_i=\left(1+\frac{4B^2}{(z-\delta^{-1})^2}\right),
\end{align}
\begin{align}
\Delta_j=\left(1+\frac{2B^2}{(z-\delta^{-1})^2}\right),
\end{align}
\begin{align}
K_{ij}=\left(1+\frac{4B^2}{(z-\delta^{-1})^2}\right),
\end{align}
where $A=\alpha_{ij}, B=(1/\delta E)^{2/3}E, C=\alpha_{ij}(1/\delta E)^{2/3}/2$ and $D=2\delta^{-1/3}E^{2/3}$, then the self-energy of
\begin{align}
\mathcal{H}^{(2*)}+V=h_i\sigma^x_i+h_j\sigma_j^x+\Delta_i\sigma_i^y+\Delta_j\sigma_j^y+\sigma^x_i\sigma^x_j+K_{ij}\sigma_i^y\sigma_j^y+V \end{align}
can be simplified as
\begin{align}
\Sigma_{-}(0)=\widetilde{\mathcal{H}}^{(2*)}+\alpha_{ij}\sigma_i^x\sigma_j^y+\mathcal{O}(\delta^{3}),
\end{align}
where $\widetilde{\mathcal{H}}^{(2*)}=\mathcal{H}^{(2*)}+\frac{B^2}{(z-\delta^{-1})^2}(\sigma_j^y+I)\mathcal{H}^{(2*)}(\sigma_j^y+I)$. Since $\Sigma_-(0)$ is a decent approximation of $\mathcal{H}^{(2*)}+V$ in the low energy subspace and the ground state of $\Sigma_-(0)$ is extremely close to that of $\widetilde{\mathcal{H}}^{(2*)}+\alpha_{ij}\sigma_i^x\sigma_j^y$, then the ground state of $\alpha_{ij}\sigma_i^x\sigma_j^y$ is extremely close to that of Hamiltonian $(\mathcal{H}^{(2*)}+V-\widetilde{\mathcal{H}}^{(2*)})$ which is only composed by $\sigma^x, \sigma^y, \sigma^x\sigma^x$ and $\sigma^y\sigma^y$.

The Hamiltonian $\sigma^y\sigma^z$ can be approximated in a similar way.


\textbf{Theorem 4.} \emph{Given an arbitrary simplified 2-local $2^N\times2^N$ Hamiltonian $\mathcal{H}=\sum_j\alpha_jP_j$ in the form of Eq.(\ref{Eq:simplified H}), where $P_j\in\{v_i^t,v_s^tv_k^t\}, t\in\{x,y,z\}$ and $\overrightarrow{\bm{\alpha}}=(\alpha_1,\alpha_2,...)$, suppose $E_0$ is the ground state energy of $\mathcal{H}$ and $|\psi_k\rangle$ is the $k$-th excited state of $\mathcal{H}$, then for a small positive value $\delta$ which is less than the spectral gap of $\mathcal{H}$, there exists a 2L-QRBM whose trial state $|\Psi(\bm{\theta}^*)\rangle$ is $\mathcal{O}(\epsilon)$ close to $\mathcal{H}$'s ground state $|\psi_0\rangle$, where $\epsilon$ is the error rate, $\tau=\mathcal{O}\left((\log1/\epsilon+N)\right)$, the Boltzmann parameter $\bm{\theta}^*=\{0, \ln(e^{\lambda^*\tau/N}+\sqrt{e^{2\lambda^*\tau/N}-1}), -\tau f(\overrightarrow{\bm{\alpha}})\}$ with $f(\alpha_j)=\alpha_j-(E_0+\delta)\sum_k \mathrm{Tr}(P_j|\psi_k\rangle\langle\psi_k|)/2^N$, which can be efficiently determined with a selected `phase shift' $\lambda^*>0$.
}


\textbf{Proof:} Since the Hamiltonian $\mathcal{H}$ is a Hermite matrix , it can be expressed as
\begin{align}
\mathcal{H}=\sum\limits_{j=0}^{2^N-1}E_j|\psi_j\rangle\langle\psi_j|,
\end{align}
where $E_j$ is the $j$-th eigenvalue whose corresponding eigenvector is $|\psi_j\rangle$. Without loss of generality, suppose the eigenvalues $\{E_j\}$ is in an increasing sequence, that is, $E_0<E_1<...<E_{2^N-1}$, then $E_0$ represents the ground state energy and $|\psi_0\rangle$ represents the ground state of the Hamiltonian $\mathcal{H}$. According to Eq.(\ref{Eq:simplified H}), for a small positive $\delta$ which is small than the spectral gap of $\mathcal{H}$, the Hamiltonian $\widetilde{\mathcal{H}}=\mathcal{H}-(E_0+\delta)\cdot I$ has the same eigenvectors as $\mathcal{H}$ and its eignvalues $\widetilde{E_j}=E_j-(E_0+\delta)$ are all non-negative. Clearly, the Hamiltonian $\widetilde{\mathcal{H}}$ can be written as
$\widetilde{\mathcal{H}}=\sum_jf(\alpha_j)P_j$, where $f(\alpha_j)=\alpha_j-(E_0+\delta)\sum_k \mathrm{Tr}(P_j|\psi_k\rangle\langle\psi_k|)/2^N$.

We first introduce the `phase shift' value $\lambda^*$ which should satisfy $\widetilde{E_0}<\lambda^{*}\leq \widetilde{E_1}$. This value can be estimated in advance by for example the phase estimation algorithm or VQE algorithm.
Let
\begin{align}
\mathcal{H}^*=\widetilde{\mathcal{H}}-\lambda^{*}I=\sum\limits_{j=0}^{2^N-1}(\widetilde{E_j}-\lambda^{*})|\psi_j\rangle\langle\psi_j|.
\end{align}
and $E_j^*=\widetilde{E_j}-\lambda^{*}$, then we have $E_0^*<0$, $E_j^*\geq0$ for $ j=1,...,2^N-1$, and the `phase shift' $\lambda^*>0$.

We now compute the ground state of $\mathcal{H}^*$ with the 2L-QRBM model. Consider $M=1$, $\bm{\theta}^*=\{m_j,W_{i1},b_i^t, K_{sk}^t\}=\{0, W_{i1}, -\tau f(\overrightarrow{\bm{\alpha}})\}$ and $\tau=\mathcal{O}\left((\log1/\epsilon+N)\right)$, then according to Eq.(\ref{Eq:2LQRBM}), the trial state of it can be expressed as
\begin{align}
|\Psi_v(\bm{\theta}^*)\rangle=c^{-1/2}\langle+|_h\exp(\mathcal{H}_{RBM}(\bm{\theta}^*))|+\rangle_h|+\rangle^{\otimes N}_v.
\end{align}
Note that $||-\tau f(\overrightarrow{\bm{\alpha}})||=\mathcal{O}(1)$, according to the Trotter theorem \cite{Nielsen2002Quantum}, the trial state can be approximated as
\begin{eqnarray}
\begin{split}
|\Psi_v(\bm{\theta}^*)\rangle=\frac{c^{-1/2}}{2^N}&\overbrace{\left(\exp(\mathcal{H}_{s}(-\tau f(\overrightarrow{\bm{\alpha}})/2T))\right)...\left(\exp(\mathcal{H}_{s}(-\tau f(\overrightarrow{\bm{\alpha}})/2T))\right)}^T\prod_{i=1}^N(e^{W_{i,1}}+e^{-W_{i,1}})I\\
&\overbrace{\left(\exp(\mathcal{H}_{s}(-\tau f(\overrightarrow{\bm{\alpha}})/2T))\right)...\left(\exp(\mathcal{H}_{s}(-\tau f(\overrightarrow{\bm{\alpha}})/2T))\right)}^T|+\rangle_v^{\otimes N}+\mathcal{O}(1/T^2).
\end{split}
\end{eqnarray}
\begin{align}
\approx\frac{c^{-1/2}}{2^N}\prod_{i=1}^N(e^{W_{i,1}}+e^{-W_{i,1}})I\exp(\mathcal{H}_{s}(-\tau f(\overrightarrow{\bm{\alpha}})))|+\rangle_v^{\otimes N},
\end{align}
where the constant $T=\mathcal{O}(\textrm{poly}(N))$.
Let $W_{i1}=\ln(e^{\lambda^*\tau/N}+\sqrt{e^{2\lambda^*\tau/N}-1})$ and note that $\mathcal{H}_{s}(-\tau f(\bm{\overrightarrow{\bm{\alpha}}}))=-\tau\widetilde{\mathcal{H}}$ , then we have
\begin{align}
|\Psi_v(\bm{\theta}^*)\rangle=c^{-1/2}\exp\left(\lambda^*\tau I\right)\exp\left(-\tau\widetilde{\mathcal{H}}\right)|+\rangle_v^{\otimes N}=c^{-1/2}e^{-\tau(\widetilde{\mathcal{H}}-\lambda^*I)}|+\rangle_v^{\otimes N}.
\end{align}
Suppose the initial state $|+\rangle_v^{\otimes N}$ of the 2L-QRBM has the overlap $K$ with the ground state, then
\begin{align}
|\Psi_v(\bm{\theta}^*)\rangle=\frac{Ke^{-(\widetilde{E_0}-\lambda^*)\tau}}{\sqrt{c}}|\psi_0\rangle&+\sqrt{1-\frac{K^2e^{-2(\widetilde{E_0-}\lambda^*)\tau}}{c}}|\psi_0^{\perp}\rangle.
\end{align}
The state $|\psi_0^{\perp}\rangle$ is composed by the eigienstates $|\psi_1\rangle,...|\psi_{2^N-1}\rangle$ that are orthogonal to $|\psi_0\rangle$, that is,
\begin{align}
|\psi_0^{\perp}\rangle=\sum\limits_{j=1}^{2^N-1}a_je^{-(\widetilde{E_j}-\lambda^*)\tau}|\psi_j\rangle,
\end{align}
where $a_j$ are the complex parameters.
Since $\widetilde{E_0}-\lambda^*<0$ and $\widetilde{E_j}-\lambda^*\geq0, j=1,...,2^N-1$, the amplitudes $a_je^{-(\widetilde{E_j}-\lambda^*)\tau} (j\geq1)$ will converge to $0$ rapidly with the increase of $\tau$, and meanwhile $\frac{Ke^{-(\widetilde{E_0}-\lambda^*)\tau}}{\sqrt{c}}$ will increase to nearly $1$ \cite{Lehto2007QITE}. Then the overlap (fidelity) between 2L-QRBM $|\Psi_v(\bm{\theta})\rangle$ and the ground state $|\psi_0\rangle$ can be estimated by
\begin{align}
F(|\Psi_v(\bm{\theta}^*)\rangle,|\psi_0\rangle)&=\frac{K^2e^{-2(\widetilde{E_0}-\lambda^*)\tau}}{K^2e^{-2(\widetilde{E_0}-\lambda^*)\tau}+\sum_j\|a_j\|^2e^{-2(\widetilde{E_j}-\lambda^*)\tau}}\\
&=1-\frac{\epsilon}{e^{-2(\widetilde{E_0}-\lambda^*)\tau}+\epsilon},
\end{align}
where $\epsilon$ represents the truncated terms $\sum_j\|a_j\|^2e^{-2(\widetilde{E_j}-\lambda^*)\tau}$. When the parameter
\begin{align}
\tau=\mathcal{O}\left((\log(1/\epsilon)+N)/\min\{\widetilde{E_j}-\lambda^*\}\right),
\end{align}
the truncated terms will converge to $0$.

\section*{Acknowledgments}
This work is supported by NSFC (Grant Nos. 61976024, 61972048, 61902166), and the Fundamental Research Funds for the Central Universities (Grant No.2019XD-A01).
\section*{Author contributions}

All authors contributed extensively to the work presented in this paper.
%
\end{document}